\documentclass[%
 prl,%
 amsmath,amssymb,
 reprint,%
 superscriptaddress
,floatfix
]{revtex4-1}

\usepackage{graphicx}
\usepackage{siunitx}
\usepackage[utf8]{inputenc}
\usepackage{amsmath}
\usepackage{amsfonts}
\usepackage{amssymb}
\usepackage{natbib}
\usepackage{xcolor}
\usepackage{graphicx} 
\graphicspath{{figures/}}
\makeatletter
\newcommand{\manualsublabel}[3]{(#2)\def\@currentlabel{\ref{#3}(#2)}\label{#1}}
\DeclareUnicodeCharacter{FB01}{}
\makeatother

\begin{document}

\title{Efficient orthogonal control of tunnel couplings in a quantum dot array}
\date{\today}

\author{T.-K. Hsiao}
\affiliation{QuTech and Kavli Institute of Nanoscience, Delft University of Technology, 2600 GA Delft, The Netherlands}
\author{C. J. van Diepen}
\affiliation{QuTech and Kavli Institute of Nanoscience, Delft University of Technology, 2600 GA Delft, The Netherlands}
\author{U. Mukhopadhyay}
\affiliation{QuTech and Kavli Institute of Nanoscience, Delft University of Technology, 2600 GA Delft, The Netherlands}
\author{C. Reichl}
\affiliation{Solid State Physics Laboratory, ETH Z\"{u}rich, Z\"{u}rich 8093, Switzerland}
\author{W. Wegscheider}
\affiliation{Solid State Physics Laboratory, ETH Z\"{u}rich, Z\"{u}rich 8093, Switzerland}
\author{L. M. K. Vandersypen}
\affiliation{QuTech and Kavli Institute of Nanoscience, Delft University of Technology, 2600 GA Delft, The Netherlands}

\begin{abstract}

Electrostatically-defined semiconductor quantum dot arrays offer a promising platform for quantum computation and quantum simulation. However, crosstalk of gate voltages to dot potentials and inter-dot tunnel couplings complicates the tuning of the device parameters. To date, crosstalk to the dot potentials is routinely and efficiently compensated using so-called virtual gates, which are specific linear combinations of physical gate voltages. However, due to exponential dependence of tunnel couplings on gate voltages, crosstalk to the tunnel barriers is currently compensated through a slow iterative process. In this work, we show that the crosstalk on tunnel barriers can be efficiently characterized and compensated for, using the fact that the same exponential dependence applies to all gates. We demonstrate efficient calibration of crosstalk in a quadruple quantum dot array and define a set of virtual barrier gates, with which we show orthogonal control of all inter-dot tunnel couplings. Our method marks a key step forward in the scalability of the tuning process of large-scale quantum dot arrays.
\end{abstract}

\maketitle

Electrostatically-defined semiconductor quantum dot arrays have great application potential in quantum computation~\citep{Loss1997,RevModPhys.79.1217,Zwa13,Vandersypen2019a} and quantum simulation~\citep{Barthelemy2013}. In these arrays, the electrochemical potentials of dots and the tunnel coupling between neighboring dots are controlled electrostatically by applying gate voltages. By adjusting the dot potentials and tunnel couplings, also the exchange coupling between electron spins in the quantum dots can be tuned to perform spin-qubit operations~\citep{Wat18,Zajac2018,Petta30092005,Laird2010}. In addition, the in-situ control of the parameters have allowed the use of quantum dot arrays for analog quantum simulation of Fermi-Hubbard physics~\citep{Hen17,Dehollain2019}. 
 
Crosstalk from capacitive coupling between gates and the quantum dot array causes a change in any of the gate voltages to affect not just one but multiple parameters. In the past, this crosstalk has been compensated through iterative adjustment of gate voltages to reach the target values. More recently, virtual gates have been introduced as linear combinations of physical gate voltages that enable orthogonal control of dot potentials~\citep{Nowack2011,Hen17}. The virtual gates are obtained by inverting a crosstalk matrix that expresses by how much each physical gate shifts each of the electrochemical potentials. The technique of crosstalk compensation for dot potentials has become a standard and essential technique in multi-dot experiments~\citep{Volk2019,Mills2019,Kandel2019}. However, the inter-dot tunnel coupling is approximately an exponential function of the gate voltages~\citep{Reed2016,Hen17,Bhattacharya1982}, and so far it has remained unclear how to incorporate this nonlinear dependence into the crosstalk matrix. Therefore, tuning of multiple tunnel couplings in a multi-dot device is mostly done by iteratively adjusting gate voltages using manual or computer-automated procedures~\citep{VanDiepen2018,Mills2019a}.

In this work, we achieve efficient orthogonal control of inter-dot tunnel couplings in a semiconductor quantum dot array. While the dependence of tunnel coupling on gate voltages is exponential, the exponent is still a linear combination of gate voltages. This allows us to extend the virtual gate matrix to include crosstalk on the tunnel barriers. Specifically, we first show how to efficiently obtain the elements of the virtual gate matrix from the derivatives of tunnel couplings with respect to gate voltages. Next, we test the use of the re-defined virtual barrier gates for orthogonal control of the tunnel couplings in a quadruple dot over a wide range of tunnel coupling values.

\begin{figure} 
\centering    
\includegraphics[width=\columnwidth]{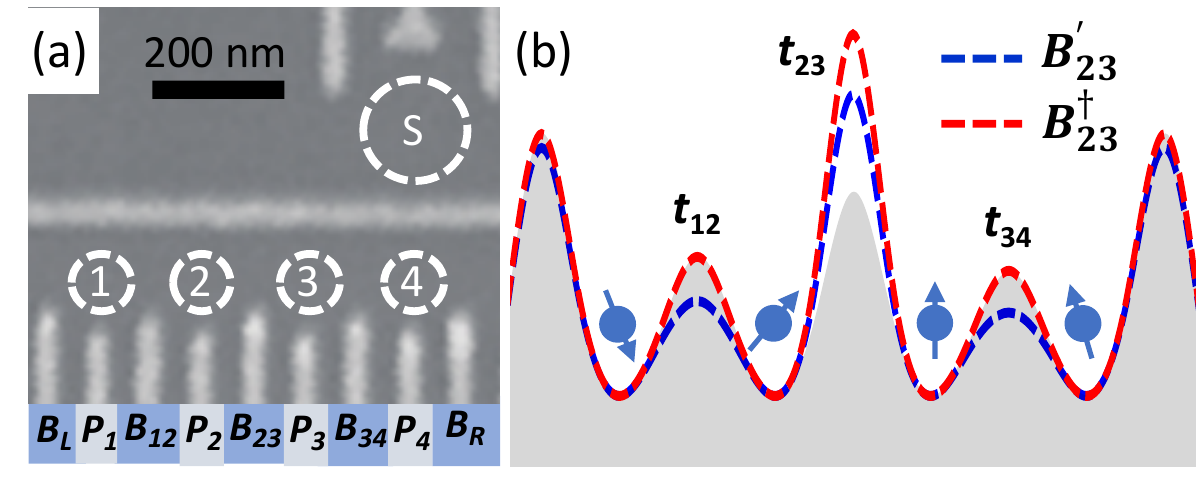}
\caption{(a) A scanning electron microscope image of a device nominally identical to the one used here. The dashed circles indicate the intended positions of the quadruple dot and sensing dot. (b) Schematics illustrating the influence of changes in $B'_{23}$ and $B^{\dagger}_{23}$ on the potential landscape of a quadruple quantum dot. The grey area denotes the original landscape, and the blue (red) dashed line indicates the landscape when $B'_{23}$ ($B^{\dagger}_{23}$) is changed. $B'_{23}$ controls the inter-dot tunnel coupling, $t_{23}$, while keeping the dot potentials fixed, but also influences $t_{12}$ and $t_{34}$. In contrast, $B^{\dagger}_{23}$ controls $t_{23}$ while affecting neither other tunnel couplings nor dot potentials.}
\label{Device_fig}
\end{figure}

The experiment is carried out in an electrostatically-defined quantum dot array in a GaAs heterostructure (see Fig.~\ref{Device_fig}(a) shows the relevant part of the device). Details of the fabrication and characterization of a nominally identical device are described in~\citep{Volk2019}. Quantum dots are formed by applying DC voltages to a set of plunger gates, $P$, and barrier gates, $B$. For brevity, we will also use the labels $P$ and $B$ to refer to the voltages applied to the corresponding gates. Each plunger gate, $P_i$, is designed to primarily control the electrochemical potential $\mu_i$ of dot $i$ and each barrier gate, $B_{ij}$, is designed to mainly control the inter-dot tunnel couplings, $t_{ij}$, between neighboring dots $i$ and $j$. Each $P_i$ is connected to a bias-tee for additional fast control of the dot potential using an arbitrary waveform generator. In this experiment, up to four dots (a quadruple quantum dot) are formed, see Fig.~\ref{Device_fig}(a). 
In addition, a sensing dot, $S$, is operated as a charge sensor. Due to capacitive coupling, the sensing dot potential and thus the conductance through the sensing dot depend on the number and position of the electrons in the quantum dot array~\citep{RevModPhys.79.1217}. The change in conductance is measured using radio-frequency reflectometry to achieve fast read-out of the charge configuration~\citep{Barthel2010}.

In the literature so far, the relationship between virtual plunger and barrier gates $P'$ and $B'$ and the physical plunger and barrier gates $P$ and $B$ is expressed via a crosstalk matrix of the form~\cite{Hen17,Volk2019,Mills2019} 
\begin{equation}
\begin{pmatrix}
    P'_1\\
    P'_2\\
    P'_3\\
    P'_4\\
    B'_{12}\\
    B'_{23}\\
    B'_{34}
\end{pmatrix}
=
\begin{pmatrix}
    1 & \alpha_{12} & \alpha_{13} & \alpha_{14} & \alpha_{15} & \alpha_{16} & \alpha_{17}\\
    \alpha_{21} & 1 & \alpha_{23} & \alpha_{24} & \alpha_{25} & \alpha_{26} & \alpha_{27}\\
    \alpha_{31} & \alpha_{32} & 1 & \alpha_{34} & \alpha_{35} & \alpha_{36} & \alpha_{37}\\
    \alpha_{41} & \alpha_{42} & \alpha_{43} & 1 & \alpha_{45} & \alpha_{46} & \alpha_{47}\\
    0 & 0 & 0 & 0 & 1 & 0 & 0\\
    0 & 0 & 0 & 0 & 0 & 1 & 0\\
    0 & 0 & 0 & 0 & 0 & 0 & 1
\end{pmatrix}
\begin{pmatrix}
    P_1\\
    P_2\\
    P_3\\
    P_4\\
    B_{12}\\
    B_{23}\\
    B_{34}
\end{pmatrix} \,.
\label{cc_matrix}
\end{equation}
The matrix entries are measured using $\alpha_{ij} = \frac{\partial \mu_i}{\partial P_j}/\frac{\partial \mu_i}{\partial P_i}$ and similar ratios involving the $B$ gates. By definition, then $\alpha_{ii} = 1$. 
The linear combination of $P$ and $B$ to orthogonally control the dot potentials is obtained from the inverse matrix. However,  $P'$-$B'$ do not compensate for the crosstalk on tunnel couplings, hence applying a voltage on $B'_{ij}$ not only changes $t_{ij}$ but also affects nearby tunnel couplings $t_{kl}$, as illustrated in Fig.~\ref{Device_fig}(b) (blue dashed line).

To overcome this limitation, we note that $t_{ij}$ can be approximated as an exponential function~\citep{Reed2016,Bhattacharya1982}
\begin{equation}
t_{ij} = t_0 \exp(\Phi_{ij}) = t_0 \exp\left(\sum_{k} {\Lambda^{ij}_{k}} P'_{k} + \sum_{kl} \Gamma^{ij}_{kl} B'_{kl}\right)
\label{tunnel_coupling_formula}
\end{equation}
where $\Phi_{ij}$ is the integral of $-\sqrt{2m_e(V_{ij}(x)-E)}$ ($m_e$ is the electron mass, $V_{ij}(x)$ is the potential of the barrier at a position $x$, and $E$ is the energy of the tunneling electron). Our crucial assumption, which we will verify below, is that $\Phi_{ij}$ can be expressed as a linear combination of $P'$ and $B'$ with coefficients $\Lambda$ and $\Gamma$ respectively.
A set of re-defined virtual gates, $P^{\dagger}$-$B^{\dagger}$, which includes the compensation for the crosstalk on tunnel couplings, is then constructed from

\begin{equation}
\begin{pmatrix}
    P^{\dagger}_1\\
    P^{\dagger}_2\\
    P^{\dagger}_3\\
    P^{\dagger}_4\\
    B^{\dagger}_{12}\\
    B^{\dagger}_{23}\\
    B^{\dagger}_{34}
\end{pmatrix}
=
\begin{pmatrix}
    1 & 0 & 0 & 0 & 0 & 0 & 0\\
    0 & 1 & 0 & 0 & 0 & 0 & 0\\
    0 & 0 & 1 & 0 & 0 & 0 & 0\\
    0 & 0 & 0 & 1 & 0 & 0 & 0\\
    \beta_{51} & \beta_{52} & \beta_{53} & \beta_{54} &1 & \beta_{56} &\beta_{57}\\
    \beta_{61} & \beta_{62} & \beta_{63} & \beta_{64} & \beta_{65} & 1 &\beta_{67}\\
    \beta_{71} & \beta_{72} & \beta_{73} & \beta_{74} & \beta_{75} &\beta_{76} & 1
\end{pmatrix}
\begin{pmatrix}
    P'_1\\
    P'_2\\
    P'_3\\
    P'_4\\
    B'_{12}\\
    B'_{23}\\
    B'_{34}
\end{pmatrix}
\label{cc_matrix_star}
\end{equation}
\\where $\beta_{51} = \Lambda^{12}_{1}/\Gamma^{12}_{12}$, $\beta_{52} = \Lambda^{12}_{2}/\Gamma^{12}_{12}$, $\beta_{56} = \Gamma^{12}_{23}/\Gamma^{12}_{12}$ and so on. 
The virtual barrier gates $B^{\dagger}_{ij}$ that orthogonally control $\Phi_{ij}$, and hence also $t_{ij}$ are obtained from the inverse matrix as a linear combination of $P'$ and $B'$. Since $P'$ and $B'$ maintain the dot potentials fixed, $B^{\dagger}$ thus achieve orthogonal control of tunnel couplings while maintaining the dot potentials fixed as well, as depicted in Fig.~\ref{Device_fig}(b) (red dashed line). Note that although $t_{ij}$ scales exponentially with $P'$ and $B'$, as long as the factors $\Lambda$ and $\Gamma$ remain the same, orthogonal control with $B^{\dagger}$ remains valid for any value of tunnel couplings. 

\begin{figure}[t!] 
\centering    
\includegraphics[width=\columnwidth]{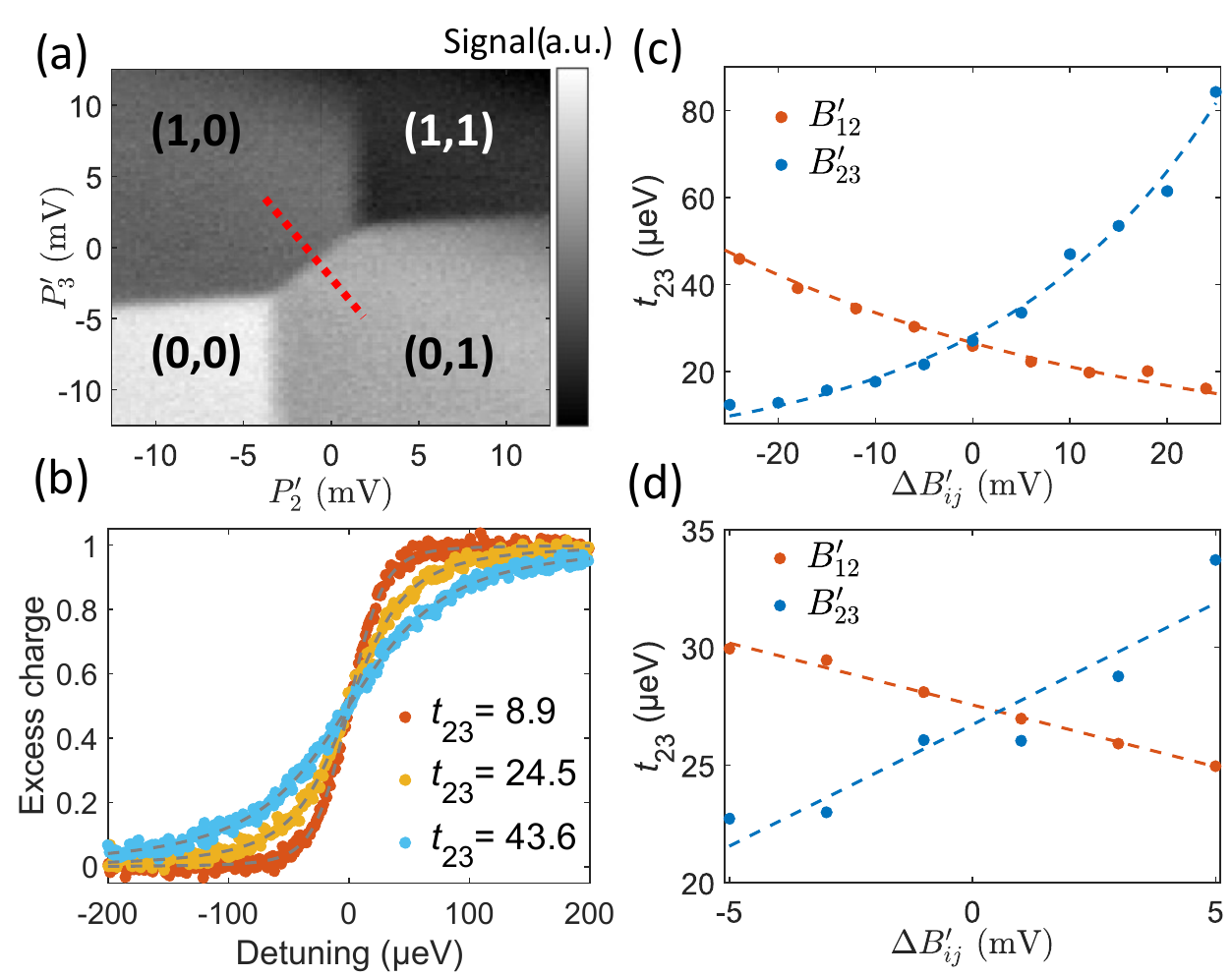}
\caption{(a) Charge stability diagram showing the sensing-dot signal as a function of voltages on $P'_2$ and $P'_3$. ($N_2$, $N_3$) indicates charge occupation of dot 2 and 3. The red dotted line indicates the inter-dot detuning axis. (b) Excess charge extracted from a fit to the sensing-dot signal as a function of detuning near the inter-dot transition in (a). Data (colored circles) for different $t_{23}$ (in \si{\micro\eV}) is shown together with the fitted curves (dashed lines). The model of the fit is described in~\citep{VanDiepen2018}. $t_{23}$ is obtained from the fit. (c) Measured tunnel coupling $t_{23}$ as a function of barrier voltage $B'_{12}$ and $B'_{23}$, with an exponential fit to the data. (d) Same as (c) but with a smaller voltage variation in $B'_{12}$ and $B'_{23}$, plotted with a linear fit.}
\label{Tunnel_coupling_measurement}
\end{figure}

We first form a double dot with dots $2$ and $3$ to illustrate how to determine $\Gamma$ from the derivatives of the tunnel couplings with respect to $B'$, see Eq.~\ref{tunnel_coupling_formula}. After the dots are formed, the crosstalk matrix from Eq.~(\ref{cc_matrix}) is determined. Figure~\ref{Tunnel_coupling_measurement}(a) shows the charge stability diagram of the double dot obtained when sweeping $P'_2$ and $P'_3$. The inter-dot tunnel coupling $t_{23}$ is characterized near the (0,1)-(1,0) inter-dot transition by scanning the dot potentials along the detuning axis (the red dotted line in Fig.~\ref{Tunnel_coupling_measurement}(a)), see Fig.~\ref{Tunnel_coupling_measurement}(b). The gate voltages are converted to dot detuning using lever arms measured with photon-assisted tunneling (PAT)~\citep{Oosterkamp1998} (see Supplemental Material~\citep{PRL_barrier_sm}). The smooth variation in charge occupation is caused by thermal excitation and charge hybridization via the inter-dot tunnel coupling, and is fitted to a model described in~\citep{VanDiepen2018}, which is adapted from the one in~\citep{DiCarlo2004}, to obtain the value of the tunnel coupling. Utilising this method, the inter-dot tunnel coupling can be measured in approximately a second. Alternatively, the tunnel coupling can also be extracted from PAT measurements~\citep{Oosterkamp1998}. The crosstalk of $B'_{kl}$ on $t_{ij}$ can be characterized by varying the voltage on $B'_{kl}$ and then measuring the change in $t_{ij}$. It is important to use the virtual barrier $B'_{kl}$ instead of the physical barrier $B_{kl}$ because varying $B'_{kl}$ keeps the dot potentials unchanged so that they remain close to the inter-dot transition. Hence, inter-dot transition scans can be performed subsequently at different $B'_{kl}$ without manually adjusting dot potentials. Note that similar methods for extracting tunnel couplings can also be used for higher electron occupations~\citep{VanDiepen2018}.

Figure~\ref{Tunnel_coupling_measurement}(c) shows the measured $t_{23}$ as a function of the corresponding barrier $B'_{23}$ and the neighboring barrier $B'_{12}$. As $B'_{23}$ becomes more positive, the potential barrier between dots $2$ and $3$ is lowered so $t_{23}$ increases exponentially. As $B'_{12}$ is increased, however, crosstalk makes $t_{23}$ decrease exponentially. The crosstalk from $B'_{12}$ to $t_{23}$ can be understood by considering the following factors.
First, increasing $B'_{12}$ also increases $B_{12}$, which by itself increases $t_{23}$.
Second, in order to keep dot potentials fixed, the voltage on $P_2$ is decreased to compensate the crosstalk from the increased voltage on $B_{12}$ to the potential of dot $2$. Decreasing $P_2$ reduces $t_{23}$. Finally, increasing $B'_{12}$ may shift the wavefunction of the electron in dot 2 away from the electron in dot 3, hence reducing the tunnel coupling as well. Combining these factors leads to the negative crosstalk of $B'_{12}$ on $t_{23}$.

By fitting the data in Fig.~\ref{Tunnel_coupling_measurement}(c) to an exponential function $t_{23}= t_0 \exp(\Gamma^{23}_{kl} B'_{kl})$, we obtain $\Gamma^{23}_{12}=-2.31\pm 0.08\times10^{-2}$\,mV$^{-1}$, $\Gamma^{23}_{23}=4.26\pm0.17\times10^{-2}$\,mV$^{-1}$ and the crosstalk ratio $r=\lvert \Gamma^{23}_{12}/\Gamma^{23}_{23} \rvert=54\pm3\%$. In fact, the ratio between $\Gamma^{23}_{12}$ and $\Gamma^{23}_{23}$ can be obtained more easily by varying $B'_{12}$ and $B'_{23}$ in a small range and measuring $\frac{\partial t_{23}}{\partial B'_{12}}$ and $\frac{\partial t_{23}}{\partial B'_{23}}$ using a linear fit (see Fig.~\ref{Tunnel_coupling_measurement}(d)). The fit gives $\frac{\partial t_{23}}{\partial B'_{12}} = -0.53\pm0.02$\,\si{\micro\eV}$/$mV, $\frac{\partial t_{23}}{\partial B'_{23}} = 1.03\pm0.18$\,\si{\micro\eV}$/$mV and the crosstalk ratio $r'=\lvert \frac{\partial t_{23}}{\partial B'_{12}}/\frac{\partial t_{23}}{\partial B'_{23}} \rvert=51\pm9\%$.
From Eq.~(\ref{tunnel_coupling_formula}), one would expect that $\Gamma^{23}_{12}/\Gamma^{23}_{23} = \frac{\partial t_{23}}{\partial B'_{12}}/\frac{\partial t_{23}}{\partial B'_{23}}$, which is confirmed by the similar ratios $r$ and $r'$ from the two different measurements in Fig.~\ref{Tunnel_coupling_measurement}(c) and (d). This result indicates that it is indeed sufficient to measure the derivatives of a tunnel coupling with respect to $B'$ to efficiently characterize the ratios between $\Gamma$, which are used for defining the $B^{\dagger}$.

Note that in this work we do not characterize the factors $\Lambda$ for $P'$ in Eq.~(\ref{tunnel_coupling_formula}). To stay near the inter-dot transition, two neighboring $P'_i$ and $P'_j$ need to be varied together, therefore $\Lambda^{ij}_i$ and $\Lambda^{ij}_j$ cannot be independently measured using our method. However, this does not affect the orthogonal control of $t_{ij}$ using $B^{\dagger}_{ij}$. In fact, the linear combination of gate voltages needed to orthogonally change $B^{\dagger}$ is independent of $\Lambda$. Of course, without knowing $\Lambda$ (here set to 0), varying $P^{\dagger}$ will affect tunnel couplings, which we return to later.

\begin{figure}[t!] 
\centering    
\includegraphics[width=\columnwidth]{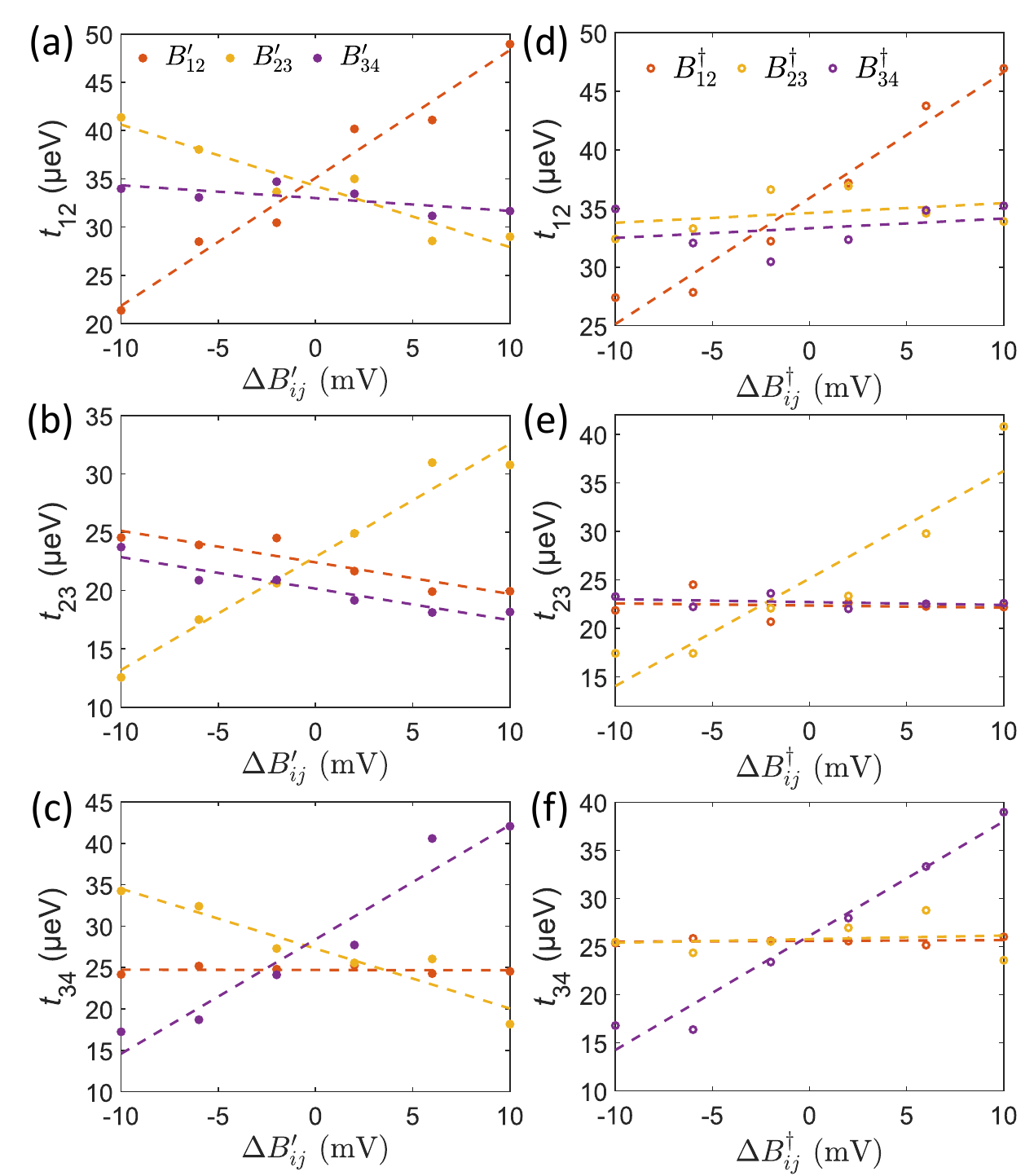}
\caption{(a-c) Measured tunnel couplings as a function of $B'$ for (a) $t_{23}$, (b) $t_{12}$, and (c) $t_{34}$. Dashed lines show linear fits to the data. (d-f) Measured tunnel couplings as a function of $B^{\dagger}$ for (d) $t_{23}$, (e) $t_{12}$, and (f) $t_{34}$. After calibration, each $t_{ij}$ only depends on the corresponding $B^{\dagger}_{ij}$. Dashed lines show linear fits to the data.}
\label{Virtual_barrier_calibration}
\end{figure}

Next, we demonstrate the crosstalk calibration and the orthogonal control of inter-dot tunnel couplings in a quadruple quantum dot, as shown in Fig.~\ref{Device_fig}(a). A quadruple dot is formed and the capacitive coupling to dot potentials is characterized for an arbitrary initial condition, where $t_{12}= 33.4\pm1.0$\,\si{\micro\eV}, $t_{23}= 23.2\pm0.4$\,\si{\micro\eV} and $t_{34}= 25.6\pm0.4$\,\si{\micro\eV}. $P'$ and $B'$ are defined with Eq.~(\ref{cc_matrix}).  The quadruple dot is then tuned to the (1,0,1,1)-(0,1,1,1) inter-dot transition to measure $t_{12}$, where ($N_1$,$N_2$,$N_3$,$N_4$) indicates the charge occupation on dots 1 to dot 4. The dependences of $t_{12}$ on $B'$ are shown in Fig.~\ref{Virtual_barrier_calibration}(a). As expected, $t_{12}$ shows the largest dependence on the corresponding barrier gate voltage $B'_{12}$.
From $\frac{\partial t_{12}}{\partial B'_{12}}=1.32\pm0.12$\,\si{\micro\eV}/mV and $t_{12}=33.4\pm1.0$\,\si{\micro\eV}, $\Gamma^{12}_{12} = 3.95\pm0.38\times10^{-2}$\,mV$^{-1}$.
Changing $B'_{23}$ has a negative crosstalk effect on $t_{12}$ ($\sim 50\%$ compared with the effect from $B'_{12}$). The crosstalk from $B'_{34}$ is weaker ($\sim 10\%$), which is expected, because $B'_{34}$ is further away from $B'_{12}$. Note that the three fitted lines roughly intersect at $\Delta B'_{ij}=0$ as expected. The deviations are caused by the error in measuring tunnel couplings. Similarly, the crosstalk on $t_{23}$ and $t_{34}$ is characterized by tuning the quadruple dot to the (1,1,0,1)-(1,0,1,1) and (1,1,1,0)-(1,1,0,1) transitions, respectively. In Fig.~\ref{Virtual_barrier_calibration}(b), $t_{23}$ shows the largest dependence on $B'_{23}$ ($\frac{\partial t_{23}}{\partial B'_{23}}=0.97\pm0.09$\,\si{\micro\eV}/mV) and $\Gamma^{23}_{23} = 4.18\pm0.39\times10^{-2}$\,mV$^{-1}$.
The crosstalk of $B'_{12}$ and $B'_{34}$ on $t_{23}$ is about $30\%$. In Fig.~\ref{Virtual_barrier_calibration}(c), $t_{34}$ shows the largest dependence on $B'_{34}$ ($\frac{\partial t_{34}}{\partial B'_{34}}=1.38\pm0.19$\,\si{\micro\eV}/mV) and $\Gamma^{34}_{34} = 5.39\pm0.51\times10^{-2}$\,mV$^{-1}$. 
The crosstalk of $B'_{23}$ on $t_{34}$ is about $50\%$ and the crosstalk of $B'_{12}$ is $<1\%$.

\begin{figure}[t!] 
\centering    
\includegraphics[width=\columnwidth]{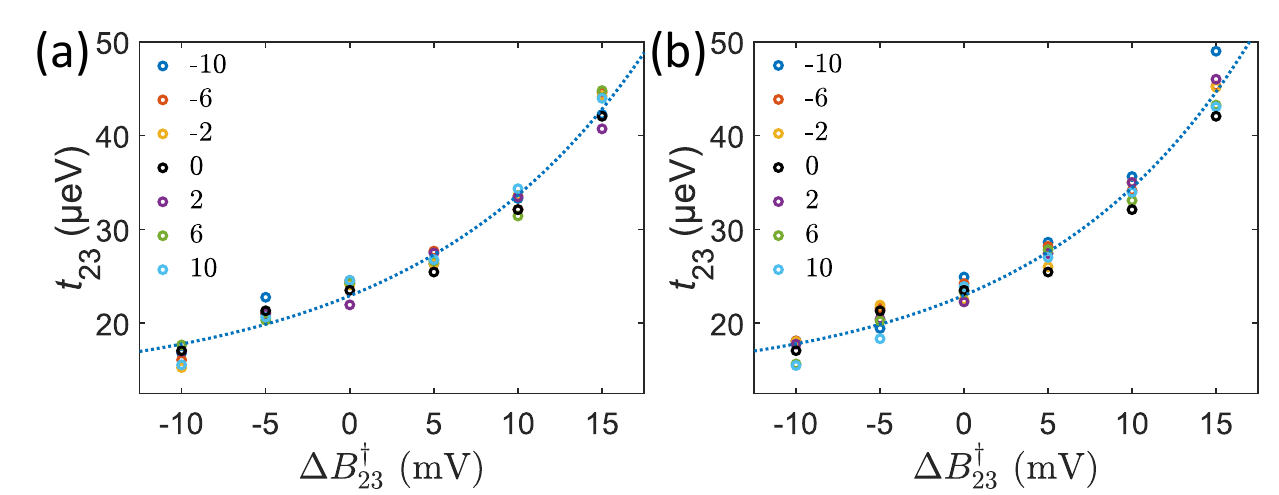}
\caption{The experimentally measured tunnel coupling $t_{23}$ as a function of $\Delta B^{\dagger}_{23}$ for different values of $\Delta B^{\dagger}_{12}$ (a) and $\Delta B^{\dagger}_{34}$ (b) (in mV), plotted with an exponential fit to the data. $\Delta B^{\dagger}_{ij}$ is the voltage relative to $B^{\dagger}_{ij}$ when $t_{ij} \sim 25$\,\si{\micro\eV}.
The exponential fit has an offset of 13\,\si{\micro\eV}. As observed in other works, the expression Eq.~\ref{tunnel_coupling_formula} is a good approximation over a finite range of gate voltages, for instance because of the presence of other tunnel barriers nearby.}
\label{2D_map}
\end{figure}

To achieve orthogonal control of the tunnel couplings, the characterized crosstalk ratios are placed into a new matrix including the tunnel-coupling crosstalk, as in Eq.~(\ref{cc_matrix_star}), and $B^{\dagger}$ are defined. If desired, the crosstalk characterization can be repeated resulting in an updated set of $B^{\dagger}$ that further reduces the residual crosstalk (see Supplemental Material~\citep{PRL_barrier_sm} for the final full matrix we used to proceed). Fig.~\ref{Virtual_barrier_calibration}(d-f) show the tunnel couplings as a function of $B^{\dagger}$. As intended, each $t_{ij}$ is only affected by the respective $B^{\dagger}_{ij}$ and crosstalk of other $B^{\dagger}$ is significantly suppressed, to $< 8\%$ for $t_{12}$ and $< 3\%$ for $t_{23}$ and $t_{34}$. The remaining crosstalk could be improved further by taking more data to accurately measure the crosstalk (see Supplemental Material~\citep{PRL_barrier_sm} for summarized crosstalk values of $B'$ and $B^{\dagger}$). This indicates that $B^{\dagger}$ orthogonally control the tunnel couplings in the quadruple dot. Using $B^{\dagger}$, we can quickly tune the quadruple dot to a desired configuration, for example, $t_{12}=t_{23}=t_{34}=33\mu$eV (see Supplemental Material~\citep{PRL_barrier_sm}).

We next verify whether $B^{\dagger}$ still compensate for crosstalk when changing the barrier gate voltages over a slightly wider range, where the exponential dependence of Eq.~(\ref{tunnel_coupling_formula}) is unmistakable. Starting from $t_{23} = 25.6\pm0.2$\,\si{\micro\eV}, the dependence of $t_{23}$ on $B^{\dagger}_{23}$ is measured for different values of $B^{\dagger}_{12}$ and $B^{\dagger}_{34}$. Fig.~\ref{2D_map}(a) and (b) show that, while changing $B^{\dagger}_{23}$ by 25\,mV exponentially increases $t_{23}$ over a range of $27$\,\si{\micro\eV}, varying $B^{\dagger}_{12}$ and $B^{\dagger}_{34}$ by 20\,mV only has a minor effect on $t_{23}$ (crosstalk $< 10\%$ except for $\Delta B^{\dagger}_{23} = -7.5$ and $-12.5$\,mV, where the small $\frac{\partial t_{23}}{\partial B^{\dagger}_{23}}$ results in a higher crosstalk ratio due to the uncertainty of the linear fit). This indicates that $B^{\dagger}$ compensate for the crosstalk in the exponent $\Phi$ rather than just compensate for the linearized the dependence of tunnel couplings in a small range of gate voltages. As long as the crosstalk coefficients $\Gamma$ for $B'$ do not change, orthogonal control of tunnel couplings using $B^{\dagger}$ is effective for a large range of tunnel coupling values.

Instead of calibrating crosstalk on all tunnel couplings in one go, we can also calibrate and compensate cross-talk one tunnel coupling at a time, as demonstrated in the Supplemental Material~\cite{PRL_barrier_sm}. This method is especially useful when some of the initial tunnel couplings are small, leading to large errors in the estimated crosstalk ratio.

Furthermore, we note that the spin exchange coupling between neighbouring spins, $J_{ij}$, is controlled by $t_{ij}$ and the double dot detuning $\epsilon_{ij}$. Since $B^{\dagger}_{ij}$ orthogonally controls $t_{ij}$ while keeping the dot potentials fixed, $B^{\dagger}_{ij}$ thus also orthogonally controls $J_{ij}$~\footnote{In the final stage of completing the manuscript, a report showing  orthogonal control of $J_{ij}$ appeared, see H. Qiao et al, arXiv:2001.02277 }. 

As mentioned earlier, we did not characterize the crosstalk factors $\Lambda$ for $P'$ since $\Lambda_i^{ij}$ and $\Lambda_j^{ij}$ cannot be independently measured using the present method. Hence, varying $P^{\dagger}$ does affect tunnel couplings. To perform the most complete crosstalk calibration, one may measure either $t_{ij}$ or $J_{ij}$ as a function inter-dot detuning, hence of $P'_i$ and $P'_j$ independently, using a spin-funnel~\cite{Petta30092005} or PAT measurement~\cite{Oosterkamp1998}. Then all the elements in the crosstalk matrix in Eq.~(\ref{cc_matrix_star}) can be obtained, allowing fully orthogonal tuning of dot potentials and tunnel couplings.

In conclusion, we have achieved orthogonal control of tunnel couplings in a quadruple dot using virtual barrier gates. The crosstalk is calibrated efficiently with a differential method, which requires only a few measurements over a small range of tunnel coupling variation. 
We also showed that the virtual barriers, calibrated at a certain condition, remain effective over a wide range of configurations. The demonstrated orthogonal control of tunnel couplings is an essential technique for configuring multi-dot devices to perform spin-qubit operations and analog quantum simulations.\\

\begin{acknowledgments}
The data reported in this paper are archived at https://doi.org/XXX~\citep{PRL_barrier_data} (url will be provided before publication)\\

We acknowledge useful discussions with members of the Vandersypen group, and technical support by O. W. B. Benningshof, N. P. Alberts, and E. van der Wiel. We also acknowledge financial support by the Dutch Research Council (NWO-Vici), the Quantera ERANET Cofund in Quantum Technologies (EU Horizon 2020), the Dutch Ministry of Economic Affairs through the allowance for Topconsortia for Knowledge and Innovation (TKI) and the Swiss National Science Foundation.
\end{acknowledgments}

\bibliography{library}

\end{document}


\title{Supplemental material for: Efficient orthogonal control of tunnel couplings in a quantum dot array}

\author{T.-K. Hsiao}
\affiliation{QuTech and Kavli Institute of Nanoscience, Delft University of Technology, 2600 GA Delft, The Netherlands}
\author{C. J. van Diepen}
\affiliation{QuTech and Kavli Institute of Nanoscience, Delft University of Technology, 2600 GA Delft, The Netherlands}
\author{U. Mukhopadhyay}
\affiliation{QuTech and Kavli Institute of Nanoscience, Delft University of Technology, 2600 GA Delft, The Netherlands}
\author{C. Reichl}
\affiliation{Solid State Physics Laboratory, ETH Z\"{u}rich, Z\"{u}rich 8093, Switzerland}
\author{W. Wegscheider}
\affiliation{Solid State Physics Laboratory, ETH Z\"{u}rich, Z\"{u}rich 8093, Switzerland}
\author{L. M. K. Vandersypen}
\affiliation{QuTech and Kavli Institute of Nanoscience, Delft University of Technology, 2600 GA Delft, The Netherlands}

\maketitle

\section{1. Determining lever arms from photon-assisted tunneling}
\begin{figure}[b!] 
\centering    
\includegraphics[width=\columnwidth]{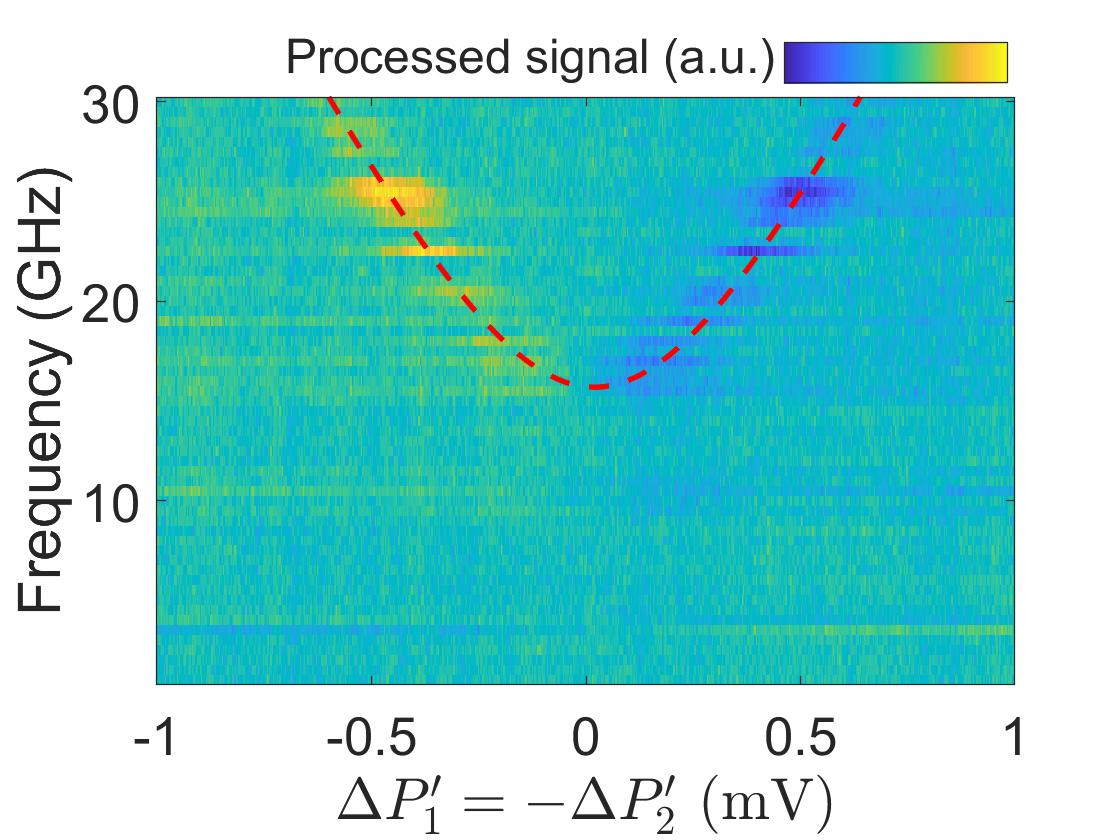}
\caption{Photon-assisted tunneling measurement showing the sensor signal as a function of frequency and detuning at the (1,0,1,1)-(0,1,1,1) inter-dot transition, shown after background subtraction~\cite{VanDiepen2018}. The red dashed line is the fit of the form $hf = \sqrt{\epsilon_{12}^2+4t_{12}^2}$.}
\label{PAT}
\end{figure}

The energy difference between states with different charge occupations can be characterized using photon-assisted tunneling (PAT)~\cite{Oosterkamp1998}, in which these states are re-populated by a resonant microwave signal. For example, at the (1,0,1,1)-(0,1,1,1) inter-dot transition and along the detuning axis where $\Delta P'_1 =- \Delta P'_2$, the energy difference between (1,0,1,1) and (0,1,1,1) is described by $hf = \sqrt{\epsilon_{12}^2+4t_{12}^2}$, where $h$ is Planck's constant, $f$ is the frequency of the microwave signal, $t_{12}$ is the inter-dot tunnel coupling, and $\epsilon_{12}$ is the detuning, which is given by $L_{12}(\Delta P'_1 - \Delta P'_2)$. $L_{12}$ is the lever arm converting gate voltage to potential energy. Fig.~\ref{PAT} shows the processed PAT signal at the (1,0,1,1)-(0,1,1,1) inter-dot transition. The fit gives $t_{12} = 31.8\pm0.5$\,\si{\micro\eV} and $L_{12} = 175\pm2$\,\si{\micro\eV}/mV. $L_{12}$ is then used for measuring $t_{12}$ from the inter-dot transition curve. By measuring the PAT signals at the (1,1,0,1)-(1,0,1,1) and (1,1,1,0)-(1,1,0,1) transitions, $L_{23} = 140\pm1$\,\si{\micro\eV}/mV and $L_{34} = 151\pm3$\,\si{\micro\eV}/mV are obtained, which are used for measuring $t_{23}$ and $t_{34}$ respectively. 

\section{2. Crosstalk values and errors}

Table~S1 summarizes the values and standard errors of crosstalk $\frac{\partial t_{ij}}{\partial B'_{kl}}$ and $\frac{\partial t_{ij}}{\partial B^{\dagger}_{kl}}$ (in \si{\micro\eV}/mV) extracted from the slopes in Fig.~3 of the main text.

\begin{table*}[t]
\centering
\begin{tabular}{ |c |c c c| c c c|} 
 \hline
  & $B'_{12}$ & $B'_{23}$ & $B'_{34}$ & $B^{\dagger}_{12}$ & $B^{\dagger}_{23}$ & $B^{\dagger}_{34}$ \\ 
 \hline
 $t_{12}$ & $1.32\pm0.12$ & $-0.63\pm0.11$ & $-0.13\pm0.12$ & $1.06\pm0.11$ & $0.08\pm0.11$ & $0.08\pm0.13$\\ [1ex]
 $t_{23}$ & $-0.27\pm0.06$ & $0.97\pm0.09$ & $-0.27\pm0.05$ & $-0.02\pm0.08$ & $1.11\pm0.22$ & $-0.03\pm0.04$\\ [1ex]
 $t_{34}$ & $0\pm0.03$ & $-0.72\pm0.12$ & $1.38\pm0.19$ & $0\pm0.02$ & $0.04\pm0.12$ & $1.19\pm0.11$\\ [1ex]
 \hline
\end{tabular}
\caption{The crosstalk $\frac{\partial t_{ij}}{\partial B'_{kl}}$ and $\frac{\partial t_{ij}}{\partial B^{\dagger}_{kl}}$ (in \si{\micro\eV}/mV) in Fig.~3. in the main text.}
\label{summarized_crosstalk}
\end{table*}

\section{3. Full crosstalk matrix}
The normalized crosstalk matrix, which defines $P^{\dagger}$-$B^{\dagger}$ in the $P$-$B$ basis, is 
{\footnotesize
\begin{equation*}
\begin{pmatrix}
    P^{\dagger}_1\\
    P^{\dagger}_2\\
    P^{\dagger}_3\\
    P^{\dagger}_4\\
    B^{\dagger}_{12}\\
    B^{\dagger}_{23}\\
    B^{\dagger}_{34}\\
\end{pmatrix}
=
\begin{pmatrix}
    1 & 0.69 & 0.32 & 0.19 & 1.38 & 0.51 & 0.20\\
    0.59 & 1 & 0.46 & 0.30 & 1.01 & 0.94 & 0.39\\
    0.23 & 0.52 & 1 & 0.39 & 0.40 & 1.12 & 0.85\\
    0.16 & 0.37 & 0.67 & 1 & 0.22 & 0.58 & 1.22\\
    0 & 0 & 0 & 0 & 1 & -0.44 & -0.03\\
    0 & 0 & 0 & 0 & -0.28 & 1 & -0.28\\
    0 & 0 & 0 & 0 & 0.06 & -0.75 & 1\\

\end{pmatrix}
\begin{pmatrix}
    P_1\\
    P_2\\
    P_3\\
    P_4\\
    B_{12}\\
    B_{23}\\
    B_{34}\\
\end{pmatrix}
\label{cc_matrix}
\end{equation*}
}
Note that, before defining $B^{\dagger}$, a set of intermediate virtual barrier gates is constructed, based on the measured crosstalk on tunnel couplings when using $B'$. The intermediate virtual barriers contain residual crosstalk on the tunnel couplings ($< 25\%$) due to measurement uncertainty. $B^{\dagger}$ are obtained by measuring and compensating for this residual crosstalk, and are then employed to show the orthogonal control of tunnel couplings in Fig.~3(d-f) in the main text.

\section{4. Tuning to a target configuration}

\begin{figure}[t!] 
\centering    
\includegraphics[width=\columnwidth]{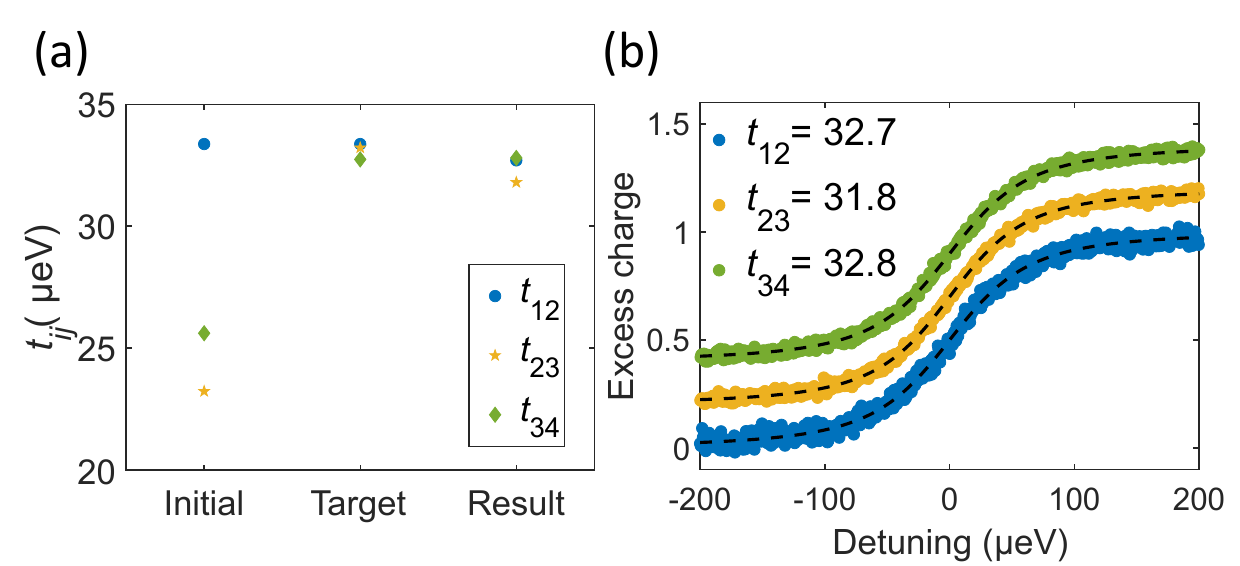}
\caption{Tuning tunnel couplings using $B^{\dagger}$. (a) The tunnel couplings in the initial configuration where $(\Delta B^{\dagger}_{12},\Delta B^{\dagger}_{23},\Delta B^{\dagger}_{34})=(0,0,0)$, the expected (`Target') and the measured (`Result') tunnel couplings in the target configuration where $(\Delta B^{\dagger}_{12},\Delta B^{\dagger}_{23},\Delta B^{\dagger}_{34})=(0,6,9)$. (b) Excess charge as a function of detuning at the $(1,0,1,1)$-$(0,1,1,1)$ (green), $(1,1,0,1)$-$(1,0,1,1)$ (yellow) and $(1,1,1,0)$-$(1,1,0,1)$ (blue) inter-dot transitions, along with the measured tunnel couplings (in \si{\micro\eV}), when $(\Delta B^{\dagger}_{12},\Delta B^{\dagger}_{23},\Delta B^{\dagger}_{34}) = (0,6,9)$. Offset in y axis is added to the data for clarity. The dashed lines show the fit to the data.}
\label{tune_using_b_dagger}
\end{figure}

Here, we demonstrate that tunnel couplings can be efficiently tuned to a target configuration using $B^{\dagger}$. As shown in Fig.~\ref{tune_using_b_dagger}(a), the initial configuration of tunnel couplings in \si{\micro\eV} is $(t_{12}, t_{23}, t_{34}) = (33.4\pm1.0,23.2\pm0.4,25.6\pm0.4)$, where $\Delta B^{\dagger}$ (in mV) are defined as $(\Delta B^{\dagger}_{12},\Delta B^{\dagger}_{23},\Delta B^{\dagger}_{34}) = (0,0,0)$.
We aim for a target configuration where all tunnel couplings $\sim 33$\,\si{\micro\eV}. According to the partial derivatives in Table~S1, applying $(\Delta B^{\dagger}_{12},\Delta B^{\dagger}_{23},\Delta B^{\dagger}_{34}) = (0,9,6)$ will in principle make $(t_{12}, t_{23}, t_{34}) = (33.4+0 \times 1.06,23.2+9 \times 1.11,25.6+6 \times 1.19) = (33.4,33.2,32.7)$.
After applying $(\Delta B^{\dagger}_{12},\Delta B^{\dagger}_{23},\Delta B^{\dagger}_{34}) = (0,9,6)$, the measurement result in Fig.~\ref{tune_using_b_dagger}(b) shows $(t_{12}, t_{23}, t_{34}) = (32.7\pm0.8,31.8\pm0.5,32.8\pm0.5)$, which is very close to the target. This result indicates that by using $B^{\dagger}$ the tunnel couplings of a quantum dot array can efficiently be tuned to a target configuration.

\section{5. Stepwise tune-and-calibrate procedure}

We here provide further details on the stepwise tune-and-calibrate procedure introduced in the main text, which allows to systematically set the tunnel couplings in a large-scale quantum dot array from an arbitrary initial configuration to a target configuration and achieving orthogonal control at the same time.

The procedure consists of the following steps:\\
1. Form the quantum dot array and define $P'$-$B'$ using the `$n+1$' method described in~\cite{Volk2019}.\\
2. Choose a $t_{ij}$, which can be chosen randomly, as the first inter-dot tunnel coupling to tune and calibrate.\\
3. Use the corresponding $B'_{ij}$ to tune $t_{ij}$ above a value ($> 20$\,\si{\micro\eV} in our case) at which the crosstalk on $t_{ij}$ can be accurately obtained with the differential method. It is preferable to directly tune $t_{ij}$ to the target value, if the target value is not too small for the differential method (otherwise see step 6).\\
4. Characterize the crosstalk of $B'$ on $t_{ij}$ and update the crosstalk matrix.\\
5. Use the updated matrix to define a new set of virtual barrier gates, $B^{\ast1}$, which compensate for the crosstalk on $t_{ij}$.\\
6. If $t_{ij}$ is not yet the target value, tune $t_{ij}$ to the target value using $B^{\ast1}_{ij}$.\\
7. Move to a $t_{kl}$ which has not been included yet in the crosstalk compensation.\\
8. Use $B^{\ast1}_{kl}$ to tune $t_{kl}$ to a sufficiently high value. Note that $t_{ij}$ will not be affected because $B^{\ast1}_{kl}$ compensates for the crosstalk on $t_{ij}$.\\
9. Characterize the crosstalk of $B^{\ast1}$ on $t_{kl}$ and update the crosstalk matrix.\\
10. Define $B^{\ast2}$, which compensate for the crosstalk on $t_{ij}$ and $t_{kl}$.\\
11. If $t_{kl}$ is not yet the target value, tune $t_{kl}$ to the target value using $B^{\ast2}_{kl}$.\\
12. Repeat steps 7--11 for the remaining tunnel couplings.\\
13. After going through all of the tunnel couplings, they are tuned to the target configuration, and the final virtual barrier gates, $B^{\dagger}$, orthogonally control the tunnel couplings.

\begin{figure}[t!] 
\centering    
\includegraphics[width=\columnwidth]{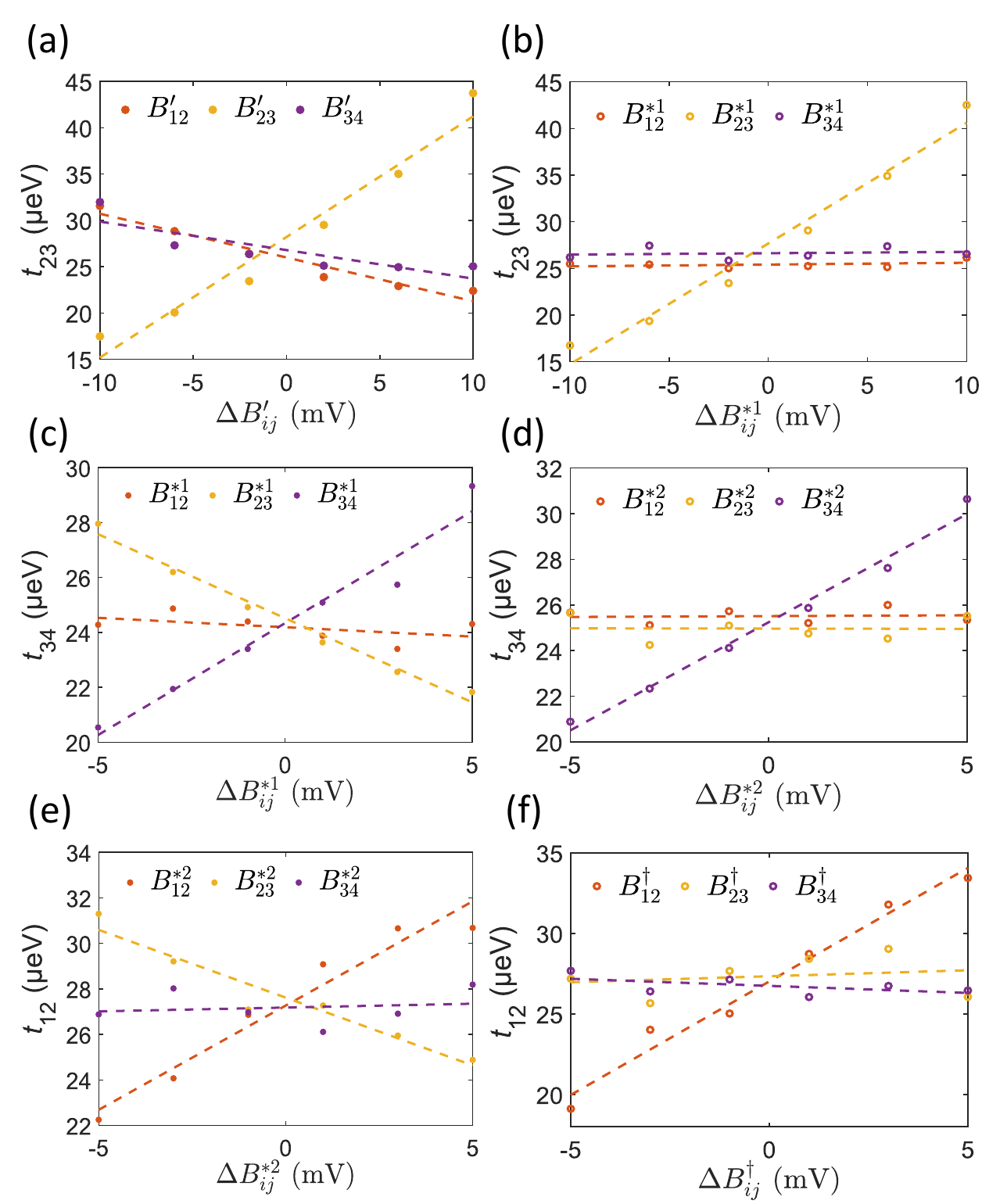}
\caption{Stepwise tuning and calibration of tunnel couplings. $t_{23}$ as a function of (a) $B'$ and (b) $B^{\ast1}$. $t_{34}$ as a function of (c) $B^{\ast1}$ and (d) $B^{\ast2}$. $t_{12}$ as a function of (e) $B^{\ast2}$ and (f) $B^{\dagger}$.}
\label{stepwise_tuning}
\end{figure}

We demonstrate this procedure on the quadruple dot to tune the tunnel couplings from an arbitrary initial configuration to a target configuration where all of the tunnel couplings $\sim 25$\,\si{\micro\eV}. The initial condition is $(t_{12}, t_{23}, t_{34}) = (6.1\pm0.4,25.9\pm0.2,8.8\pm0.4)$ \si{\micro\eV}. After $P'$-$B'$ are defined, the procedure is first carried out on $t_{23}$. Fig.~\ref{stepwise_tuning}(a) shows the crosstalk of $B'$ on $t_{23}$. Based on the characterized crosstalk, $B^{\ast1}$ are defined using a crosstalk matrix where the sub-matrix for the barrier gates is

\begin{equation*}
\begin{pmatrix}
    B^{\ast1}_{12}\\
    B^{\ast1}_{23}\\
    B^{\ast1}_{34}
\end{pmatrix}
=
\begin{pmatrix}
    1 & 0 & 0\\
    -0.36 & 1 & -0.24\\
    0 & 0 & 1\\
\end{pmatrix}
\begin{pmatrix}
    B_{12}\\
    B_{23}\\
    B_{34}
\end{pmatrix}
\label{cc_matrix_ast1}
\end{equation*}
The physical gate voltages corresponding to $B^{\ast1}$ are obtained from the inverse matrix. In Fig.~\ref{stepwise_tuning}(b), using $B^{\ast1}$, the crosstalk on $t_{23}$ is reduced to below $2\%$, showing the compensation for the crosstalk. Subsequently, $t_{34}$ is tuned to $24.7\pm0.2$\,\si{\micro\eV} using $B^{\ast1}_{34}$ ($\Delta B^{\ast1}_{34}=105$\,mV). Interestingly, since $B^{\ast1}_{34}$ includes the compensation for crosstalk on $t_{23}$, changing $B^{\ast1}_{34}$ by 105\,mV only affects $t_{23}$ by $0.7$\,\si{\micro\eV} (from $25.9\pm0.2$\,\si{\micro\eV} to $26.6\pm0.3$\,\si{\micro\eV}). This shows that $t_{34}$ can be tuned using $B^{\ast1}_{34}$ without disturbing $t_{23}$. The crosstalk of $B^{\ast1}$ on $t_{34}$ is shown in Fig.~\ref{stepwise_tuning}(c). The crosstalk matrix is updated by multiplying the matrix describing the crosstalk of $B^{\ast1}$ on $t_{34}$ by the current matrix used for defining $B^{\ast1}$, and then normalizing each row so that the diagonal elements are $1$. The updated virtual barrier gates $B^{\ast2}$, which compensate for the crosstalk on $t_{23}$ and $t_{34}$, are defined as

\begin{equation*}
\begin{pmatrix}
    B^{\ast2}_{12}\\
    B^{\ast2}_{23}\\
    B^{\ast2}_{34}
\end{pmatrix}
=
\begin{pmatrix}
    1 & 0 & 0\\
    -0.36 & 1 & -0.24\\
    0.23 & -0.64 & 1\\
\end{pmatrix}
\begin{pmatrix}
    B_{12}\\
    B_{23}\\
    B_{34}
\end{pmatrix}
\label{cc_matrix_ast2}
\end{equation*}
In Fig.~\ref{stepwise_tuning}(d), using $B^{\ast2}$, the crosstalk on $t_{34}$ is suppressed to below $1\%$. Next, $t_{12}$ is tuned to $27.7\pm0.6$\,\si{\micro\eV} using $B^{\ast2}_{12}$ ($\Delta B^{\ast2}_{12}=100$\,mV). Again, since $B^{\ast2}_{12}$ includes the compensation for crosstalk on $t_{23}$ as well, changing $B^{\ast2}_{12}$ by 100\,mV only affects $t_{23}$ by $2.4$\,\si{\micro\eV} (from $26.6\pm0.3$\,\si{\micro\eV} to $24.2\pm0.2$\,\si{\micro\eV}). Repeating the crosstalk characterization on $t_{12}$ in Fig.~\ref{stepwise_tuning}(e), $B^{\dagger}$, which include compensation for the crosstalk on all the tunnel couplings, are defined as

\begin{equation*}
\begin{pmatrix}
    B^{\dagger}_{12}\\
    B^{\dagger}_{23}\\
    B^{\dagger}_{34}
\end{pmatrix}
=
\begin{pmatrix}
    1 & -0.84 & 0.20\\
    -0.36 & 1 & -0.24\\
    0.23 & -0.64 & 1\\
\end{pmatrix}
\begin{pmatrix}
    B_{12}\\
    B_{23}\\
    B_{34}
\end{pmatrix}
\label{cc_matrix_dagger}
\end{equation*}
In Fig.~\ref{stepwise_tuning}(f), using $B^{\dagger}$, the crosstalk on $t_{12}$ are reduced to below $6\%$. The tunnel couplings have been tuned from an initial configuration where $(t_{12}, t_{23}, t_{34}) = (6.1\pm0.4,25.9\pm0.2,8.8\pm0.4)$ \si{\micro\eV} to $(27.7\pm0.6,24.2\pm0.2,24.7\pm0.2)$ \si{\micro\eV}, which is close to the target $(25,25,25)$ \si{\micro\eV}. Note that the gate voltages used in Fig.~\ref{stepwise_tuning} and those in Fig.~3 are different. The different potential profiles caused by the gate voltages may explain the different crosstalk ratios in the two crosstalk matrices.

In summary, we have demonstrated the stepwise tune-and-calibrate procedure to tune the quadruple dot to a target configuration. In addition, $B^{\dagger}$ include the compensation for the crosstalk on all the tunnel couplings, so $B^{\dagger}$ can be used to orthogonally tune the tunnel couplings to other configurations provided that the crosstalk ratios remain the same. 

\bibliography{library}